\documentclass[runningheads]{llncs}
\usepackage[T1]{fontenc}
\usepackage{graphicx}
\usepackage{amssymb}
\usepackage{amsmath}
\usepackage{url}
\usepackage{multirow} 
\usepackage{rotating}
\usepackage{color}
\usepackage{algorithm}
\usepackage{algpseudocode}
\begin{document}
\title{StoDIP: Efficient 3D MRF image reconstruction with deep image priors and stochastic iterations}
\titlerunning{StoDIP: Efficient 3D MRF image reconstruction}
%
\author{Perla Mayo\inst{1}\and
Matteo Cencini\inst{2}\and
Carolin M. Pirkl\inst{3}\and
Marion I. Menzel\inst{3,4}\and
Michela Tosetti\inst{5}\and
Bjoern H. Menze\inst{6}\and
Mohammad Golbabaee\inst{1} 
}
\authorrunning{P. Mayo et al.}
%
\institute{
University of Bristol, UK \\
\email{\{pm15334,m.golbabaee\}@bristol.ac.uk} \and
INFN Pisa division, Italy\and
GE HealthCare, Munich, Germany\and
Technische Hochschule Ingolstadt, Germany\and
IRCCS Stella Maris, Italy\and
University of Zurich, Switzerland
}
\maketitle              

\begin{abstract}
Magnetic Resonance Fingerprinting (MRF) is a time-efficient approach to quantitative MRI for multiparametric tissue mapping. The reconstruction of quantitative maps requires tailored algorithms for removing aliasing artefacts from the compressed sampled MRF acquisitions. Within approaches found in the literature, many focus solely on two-dimensional (2D) image reconstruction, neglecting the extension to volumetric (3D) scans despite their higher relevance and clinical value. A reason for this is that transitioning to 3D imaging without appropriate mitigations presents significant challenges, including increased computational cost and storage requirements, and the need for large amount of ground-truth (artefact-free) data for training. To address these issues, we introduce StoDIP, a new algorithm that extends the ground-truth-free Deep Image Prior (DIP) reconstruction to 3D MRF imaging. StoDIP employs memory-efficient stochastic updates across the multicoil MRF data, a carefully selected neural network architecture, as well as faster nonuniform FFT (NUFFT) transformations. This enables a faster convergence compared against a conventional DIP implementation without these features. Tested on a dataset of whole-brain scans from healthy volunteers, StoDIP demonstrated superior performance over the ground-truth-free reconstruction baselines, both quantitatively and qualitatively.
\end{abstract}

\section{Introduction}
Magnetic Resonance Fingerprinting (MRF)\cite{ma2013mrf} is an advanced imaging technique within quantitative MRI that enables the simultaneous measurement of multiple tissue parameters, such as T1 and T2 relaxation times, through time-efficient scans. Unlike the conventional qualitative MRI approach, where weighted images are influenced by both tissue properties and scan parameters, MRF produces repeatable and reproducible mapping of tissue quantitative parameters, allowing for objective comparisons across a range of clinical applications \cite{hsieh2020mrfclinical,poorman2020mrfclinical}.

MRF data has a spatiotemporal structure: transient-state time signals (fingerprints) that encode tissue properties are sampled across the k-space i.e. spatial Fourier domain. Compressed sensing is applied to accelerate the (otherwise prohibitive) acquisitions through subsampling only a fraction of the k-space information, however, shorter scan-times introduce undesired image aliasing artefacts. To address this issue, image reconstruction algorithms ranging from non-iterative dictionary matching~\cite{ma2013mrf,mcgivney2014svdmrf}, to iterative compressed sensing variations ~\cite{davies2014csmrf,zhao2016mlemrf,wang2016mrfcsdistancemetriclearning,asslander2018admm,zhao2018lowrank}
and more recently, deep learning~\cite{fang2019supervisedmrf,hamilton2022dipmrf,chen2020blochnet,chen20203mrfsupervised,fang2019rcaunetmrf} have been proposed. While deep learning holds great promise for MRF image reconstruction, the current reliance on supervised learning methods on datasets containing clean, artifact-free "ground-truth" quantitative maps for training remains a practical challenge. This is because obtaining such datasets is difficult as it would involve dense acquisitions with long, clinically infeasible scan times, susceptible to motion artefacts, among other factors.

Deep Image Prior (DIP)~\cite{ulyanov2018deiporiginal} has emerged as a successful unsupervised approach for ground-truth-free imaging with applications ranging from PET imaging \cite{gong2018petdip} to  compressed sensing MRI~\cite{yoo2021dip2dheart}, and recently, MRF~\cite{hamilton2022dipmrf}. DIP utilises the well-crafted architecture of convolutional neural networks as an implicit image prior without the need for pretraining them on a dataset. Additionally, DIP incorporates the knowledge of the physical acquisition model through iteratively minimising a data consistency loss e.g., in MRI or MRF, a loss on differences between the measured and predicted (from the reconstruction) subsampled k-space data. The downside of DIP is its high computational demand, requiring a large number of iterations with slow and sometimes unstable convergence. DIP approaches are currently applicable/examined only in 2D MRF settings despite the higher relevance and clinical value of 3D volumetric data. Transition to 3D imaging, without appropriate mitigations, would impose a significant challenge due to the increased computational cost and storage requirements. 

\emph{Our contribution:} In this work, we present StoDIP, a DIP-based algorithm capable of processing volumetric MRF data. Our approach employs memory-efficient stochastic iterations across the multicoil scan data, carefully analysed/selected hyperparameters (including the backbone network architecture and additional regularization), and also a faster nonuniform Fast Fourier Transform (NUFFT) based on the newly released cuFINUFFT library~\cite{shih2021cufinufft}. This results in the faster convergence of a DIP-based model, as opposed to implementations without these features. Methods are evaluated on a dataset of healthy volunteers' whole-brain scans, accelerating the current 8-minute acquisition time retrospectively by twice. StoDIP shows competitive performance against the tested ground-truth-free reconstruction baselines quantitatively and qualitatively, enabling new venues for efficient 3D DIP implementations.


\section{Methods}
The MRF image reconstruction is an  inverse problem:
\begin{equation}
    \label{eq:inverse_problem}
    \textbf{y} \approx A(\textbf{x}),\;
    \textrm{such that}\; \textbf{x}_v=\text{PD}_v \cdot \mathcal{B}(\text{T1}_v,\text{T2}_v), \;\forall v : \text{voxels}, 
\end{equation}
where, $\textbf{y} \in \mathbb{C}^{C \times M\times L\times T}$ are the undersampled k-space measurements from $M$ spatial frequency locations across $L$ (e.g. spiral or radial) arms, $C$ scanner coils, and $T$ timeframes. The linear acquisition/forward operator $A: \mathbb{C}^{N \times K} \rightarrow \mathbb{C}^{C \times M\times L\times T}$ includes coil sensitivities, NUFFT and temporal Singular Value Decomposition (SVD) for dimensionality reduction~\cite{mcgivney2014svdmrf}. The aim is to reconstruct the timeseries of magnetisation images (TSMI) $\textbf{x} \in \mathbb{C}^{N \times K}$ that contain the tissue-encoding signals of length $K$ for $N$ voxels. The TSMI is commonly time-compressed i.e. $K\ll T$ to enable memory efficiency and faster computations \cite{golbabaee2021lrtv,asslander2018admm}. After reconstructing the TSMI $\textbf{x}$, the MRF dictionary-matching (DM)~\cite{ma2013mrf,mcgivney2014svdmrf} is applied as postprocessing to obtain the desired quantitative maps (Q-Maps), here the T1 and T2 relaxations times and proton density (PD). In this sense the DM process inverts the nonlinear Bloch response model $\mathcal{B}: \mathbb{C}^{N \times 3} \rightarrow \mathbb{C}^{N \times K}$ that voxelwise relates the Q-Maps to the TSMI.

\subsection{Stochastic Deep Image Prior algorithm (StoDIP)}
\label{sec:methods}
Our proposed approach (StoDIP) iteratively estimates a cleaner version of TSMI $\hat{\textbf{x}}$ from a corrupted, alias-contaminated input \(\textbf{x}^{(0)} \in \mathbb{C}^{N \times K}\). This is done through minimising the following k-space consistency loss according to~\eqref{eq:inverse_problem}:  
\begin{equation}
    \mathcal{L}(\theta) = ||\sqrt{DCF}\cdot \textbf{y}- \sqrt{DCF} \cdot A(\hat{\textbf{x}})||_2^2, \; \text{where}\; \hat{\textbf{x}}=G_{\theta}(\textbf{x}^{(0)})
    \label{eq:loss}
\end{equation}
Here $G_{\theta}$ (generator) is a variant of U-Net \cite{ronneberger2015unet} parameterised by optimisable convolutional layers $\theta$.  The density compensation function (DCF) is used as preconditioner to accelerate the optimisation~\cite{hamilton2022dipmrf}. The k-space loss $\mathcal{L}=\sum_{c=1}^{C} \mathcal{L}_c$ can be decomposed across the coil dimension, where:
\begin{equation}
    \mathcal{L}_c = ||\sqrt{DCF}\cdot \textbf{y}_c - \sqrt{DCF} \cdot A_c(\hat{\textbf{x}})||_2^2, 
    \label{eq:loss_coil}
\end{equation}
and $\textbf{y} _c$ and  $A_c$ denote the splitted k-space data and forward operator for each coil $c$, respectively i.e., $\textbf{y} =\{\textbf{y}_c\}_{c=1}^C$ and $A=\{A_c\}_{c=1}^C$. Since processing the entire k-space loss at once for 3D MRF scans is not feasible in common GPU memories, we stochastically iterate through the loss one coil at a time and update the paramaters of the network per coil iteration, as opposed to waiting for the information of all coils to be aggregated i.e. a full gradient step. We demonstrate that these stochastic updates can further speed up the training of the network. The pseudocode for StoDIP (with PyTorch syntax) is provided in Algorithm \ref{alg:3ddipmrf}. Note that to convert the stochastic updates to a full gradient step, it suffices to swap lines 2 and 3, as well as lines 8 and 9.

During our experiments, we observed that the output of StoDIP, albeit good, contained mild checkerboard artefacts. To combat this, we further regularise the model with a spatial-domain total variation penalty on $\hat{\textbf{x}}$ (StoDIP + TV) which resulted in cleaner TSMI images. This regularisation with hyperparameter $\lambda>0$ is added to the k-space loss $\mathcal{L}_c$ as line 6 in Algorithm \ref{alg:3ddipmrf}.

\begin{algorithm}[t]
    \caption{StoDIP iterations}\label{alg:3ddipmrf}
    \begin{algorithmic}[1]
        \Require $\textbf{x}^{(0)}, \textbf{y} =\{\textbf{y}_c\}_{c=1}^C, A=\{A_c\}_{c=1}^C, \text{DCF}, \text{max\_epochs}, \text{lr-scheduler}$
        \For{$\text{epoch}=1,\hdots,  \text{max\_epochs}$}
            \For{$c \in \text{rand.perm}([1,\hdots C])$}
                \State $\hat{\textbf{x}} = G(\textbf{x}^{(0)})$
                \State $\hat{\textbf{y}}_{c} = A_c(\hat{\textbf{x}}) $
                \State $\mathcal{L}_c = ||\sqrt{\text{DCF}}\cdot \textbf{y}_{c}- \sqrt{\text{DCF}} \cdot \hat{\textbf{y}}_{c}||_2^2$
                \State $\text{loss} = \mathcal{L}_c + \lambda TV(\hat{\textbf{x}})$
                \State $\text{loss.backward()}$
                \State $\text{optimiser.step()}$
            \EndFor
            \State $\text{lr-scheduler.step()}$
        \EndFor
    \end{algorithmic}
\end{algorithm}

\subsection{Hyperparameters}
\label{subsec:hyperparameters}
We investigated the impact of various training options and parameters on the model's performance, with the main factors including the selection of the NUFFT operator, the U-Net architecture used as the generator, the learning rate and its scheduler (or lack thereof), and the choice of initialiser $\textbf{x}^{(0)}$. 

\begin{center}
\begin{figure}[t]
\begin{tabular}{cc}
    \includegraphics[width=0.5\textwidth]{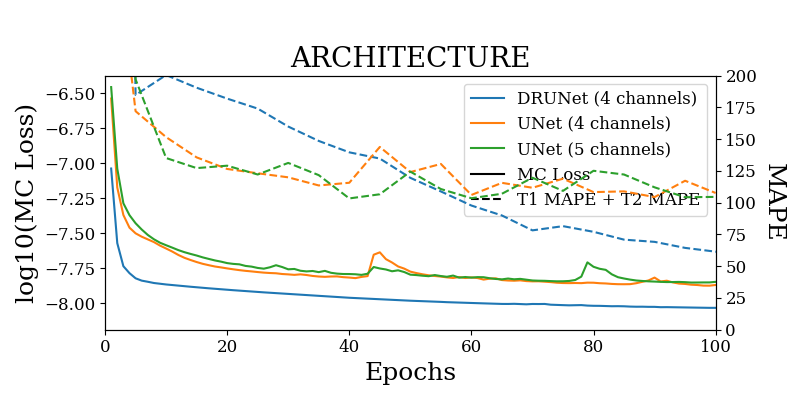} &
    \includegraphics[width=0.5\textwidth]{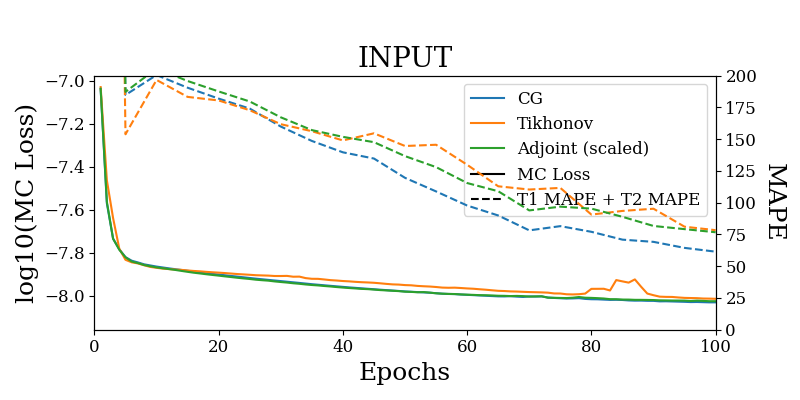} \\ 
    (a) & (b) \\
    \includegraphics[width=0.5\textwidth]{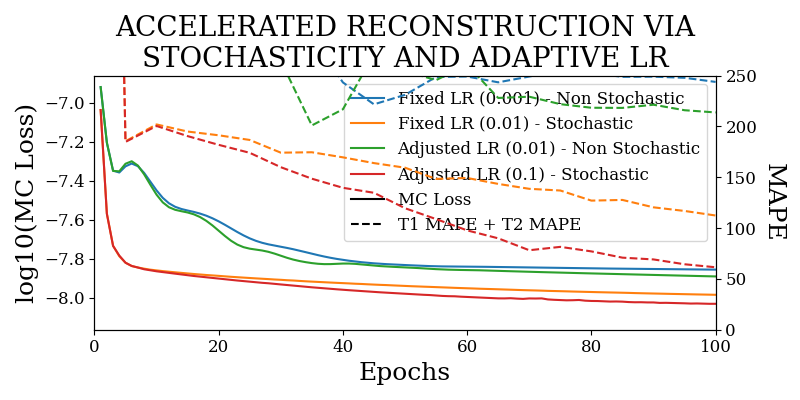} & 
    \includegraphics[width=0.5\textwidth]{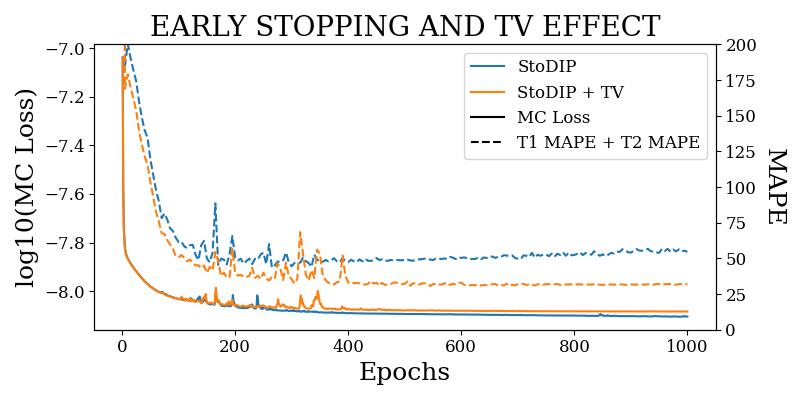}\\
    (c) & (d) \\
\end{tabular}
\caption{Model performance during training for k-space (measurement consistency, MC) loss, (solid lines) averaged across the eight coils, and the T1 + T2 MAPEs (dashed lines, using only for monitorning, not for training) for different subsets of experiments, each one showcasing a behaviour for a particular case of study: a) architecture choice, b) different \(\textbf{x}^{(0)}\), c) stochasticity and choice of LR, and d) early stopping and the effect of TV spatial regularisation. } \label{fig:mc_loss_learning_rates}
\end{figure}
\end{center}

We assessed the performance of two NUFFT libraries available on PyTorch, torchkbnufft \cite{muckley2020tkbnufft} and cuFINUFFT \cite{shih2021cufinufft}. From our assessments, we concluded that cuFINUFFT was better suited for the iterative learning of our approach, as it was capable of computing both forward and backward operations efficiently without sacrificing accuracy. Specifically, a forward-backward pass of Algorithm \ref{alg:3ddipmrf} is two orders of magnitude faster when using cuFINUFFT than the alternative. Details on other training options are provided next whilst Figures ~\ref{fig:mc_loss_learning_rates}a-d demonstrate their performance vs training iterations.

\textbf{Generator Backbone (U-Net)} We tested the original DIP architecture proposed in \cite{ulyanov2018deiporiginal} and DRUNet \cite{zhang2021drunet}, an architecture integrating residual blocks within a U-Net model. To fit these architectures in GPU memory, we performed the following adjustments. For the DIP architecture, we used the specifications for the super resolution task with four and five downsampling/upsampling blocks with [16, 32, 64, 128] and [16, 32, 64, 128, 256] feature channels, respectively. Both variations used trilinear interpolation for upsampling (the 3D extension of bilinear). For DRUNet, we extended the available 2D implementation \footnote{\url{https://github.com/cszn/DPIR}} and used the same four downsampling/upsampling units as described previously for the DIP model, along with 2 residual units, and bias enabled. Other architectures design remained as described in the original papers.

\textbf{Learning Rate}
During hyperparameter tuning, we observed that, when using a fixed learning rate, the model with the DRUNet architecture reached a loss plateau, showing minimal improvements after a couple of hundred of iterations. As this phenomenon could be attributed to reaching a saddle point in the loss landscape, we implemented a scheduler to navigate this area. This consisted in regular linear increases in the learning rate whenever there was an improvement in volume loss (averaged across coils) and decreases otherwise. Performance of both alternatives is shown in Fig. ~\ref{fig:mc_loss_learning_rates}c.

\textbf{Initialisation of $\textbf{x}^{(0)}$}
Instead of beginning with random noise like \cite{hamilton2022dipmrf}, we initialise an educated estimate \(\textbf{x}^{(0)}\) by utilising the solution to the inverse problem in Eq. \eqref{eq:inverse_problem}. We achieve this using: a) the SVD-MRF with a scaling factor \cite{mcgivney2014svdmrf}, b) low-rank optimisation with conjugate gradient, and c) low-rank optimisation with Tikhonov regularisation. We also tested random input, as suggested in DIP approaches, but failed to produce comparable results and was thus not included.

The TV hyperparameter $\lambda$ also required fine tuning. For this, we performed several evaluations starting from $\lambda=1$ to $\lambda=1e-9$, with reductions by a factor of 10 until the best performance was found. 

\vspace{-.5cm}
\section{Implementation Details}
\label{sec:implementation}
The algorithm was fully implemented in PyTorch, with additional libraries utilized to facilitate certain aspects of the pipeline. Specifically, MONAI\cite{monai} assisted in the DRUNet implementation, while the computation of the NUFFT was facilitated by the PyTorch wrapper\footnote{\url{https://flatironinstitute.github.io/pytorch-finufft/}} for cuFINUFFT \cite{shih2021cufinufft}. This enabled efficient execution on GPU and provided necessary functionality for gradient computation. The code was executed on an NVIDIA GeForce RTX 4090, which could handle both network training and NUFFT operations on a single GPU device.

From the hyperparameter exploration step described in an earlier section, we chose the following design for the generator of the final model to evaluate: DRUNet with adaptive learning rate (alternating between increases and decreases). The learning rate schedules for StoDIP consisted of linear increases from 0.001 to 0.01, and linear decreases from 0.01 to 0.001, each in a space of 250 epochs. To process the TSMI data, we formatted it from complex to real-valued by splitting the real and imaginary information into two separate channels. Consequently, the generator takes as input (and produces as output) an array of real numbers with dimensions \(N \times 2\cdot K\). Each volume was trained for 500 epochs (i.e. 4k iterations across the 8 coils), using ADAM optimiser, with $\lambda=1e-07$ in the loss function whenever TV was enforced.

\subsection{Experiments}
We benchmarked our method on accelerated k-space measurements (R=2) against the classical approaches SVD-MRF\cite{mcgivney2014svdmrf}, the low rank (LR) inspired by Zhao et al. \cite{zhao2018lowrank}, and low rank with Tikhonov regulariser. The dataset is comprised of 5 volumes of healthy individuals acquired on a GE HealthCare HDxt 1.5 T scanner using a quadrature body coil for radiofrequency transmission and an 8 channel receive only head coil for signal reception. The scans have been acquired following the QTI procedure for transient-state signals \cite{gomez2019qtimrf} consisting of an Inversion-prepared unbalanced Steady State Free Precession MR sequence with a variable flip angle train of 880 pulses and fixed Repetition Time (TR) of 10.5 ms. MR signal was spatially encoded with a 1.125 mm isotropic resolution 3D spiral trajectory, using 56 interleaves to sample each frame in the flip angle train for a total acquisition time of 8 min 37 sec. The resulting k-space data has dimensions \(8 \times 876\times 56 \times 880\), (C \(\times\) M \(\times\) L \(\times\) T), while the dimension of each reconstructed image space volume is \(200\times 200 \times 200\) voxels. To retrieve the accelerated scans, we subsampled the k-space retrospectively across the L dimension, retrieving every other arm in the array, effectively using k-space of dimensions \(8 \times 876\times \textbf{28} \times 880\). The reference value used to compute the evaluation metrics was obtained by solving Eq. \eqref{eq:inverse_problem} via LRTV\cite{golbabaee2021lrtv} on the original raw data (i.e. no acceleration).

To monitor performance progress, we mapped TSMI to Q-Maps every five epochs. To mitigate the lengthy processing times associated with Dictionary Matching, we implemented a fully connected network to directly map TSMI to Q-Maps \cite{cohen2018drone}, significantly reducing processing time. However, for the final assessment, we employed DM on all techniques at the conclusion of the training process (epoch 500). We report in Table \ref{tab:results} the Mean Average Percentage Error (MAPE) of T1/T2 maps, and Peak Signal-to-Noise Ratio (PSNR) and Structural Similarity (SSIM) Index for T1/T2 and normalised PD. To accompany these metrics we also offer in Fig. ~\ref{fig:qmaps} the reconstructed T1 and T2 maps for the different approaches.

\section{Results and Discussions}
\label{sec:discussion}

\begin{table}[t]
    \begin{center}
        \caption{Metrics for reconstructed T1 and T2 maps from accelerated k-space data (acceleration factor of 2). The best metric for each tissue map is indicated in bold.}
        \label{tab:results}
        \bgroup
        \setlength\tabcolsep{0.2cm}
        \def\arraystretch{1.1}
        \begin{tabular}{|l||c|c||c|c|c||c|c|c| }
            \hline
            \multirow{ 2}{*}{Approach} & \multicolumn{2}{c||}{MAPE (\%)} & \multicolumn{3}{c||}{PSNR (dB)} & \multicolumn{3}{c|}{SSIM}  \\ \cline{2-9}
                           & T1 & T2 & T1 & T2 & PD & T1 & T2 & PD  \\
            \hline \hline
           LR  
                & 19.38 & 48.83 & 31.78 & 29.05 & 26.77 & 0.96 & 0.93 & 0.90 \\
            SVD-MRF
                & 13.75 & 68.40 & 33.18 & 26.53 & 28.30 & 0.97 & 0.91 & 0.92 \\
            LR-Tikh   
                & \textbf{8.98} & 28.54 & \textbf{37.33} & 31.47 & \textbf{32.00} & \textbf{0.98} & 0.95 & 0.93 \\ \hline
            StoDIP
                & 10.59 & 23.78 & 35.00 & 34.08 & 31.70 & 0.98 & 0.97 & 0.93 \\
            StoDIP + TV
                & 9.08 & \textbf{16.04} & 35.25 & \textbf{38.52} & 30.74 & \textbf{0.98} & \textbf{0.98} & \textbf{0.94} \\
            \hline
        \end{tabular}
        \egroup
    \end{center}
\end{table}

\begin{figure}[t]
    \begin{tabular}{cc}
        T1 & T2 \\
        \includegraphics[width=0.24\textwidth]{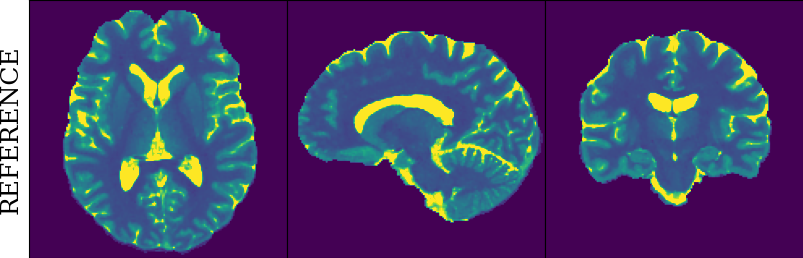} 
        \includegraphics[width=0.23\textwidth]{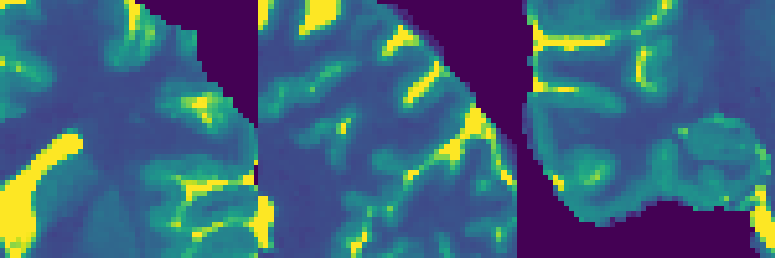} &
        \includegraphics[width=0.23\textwidth]{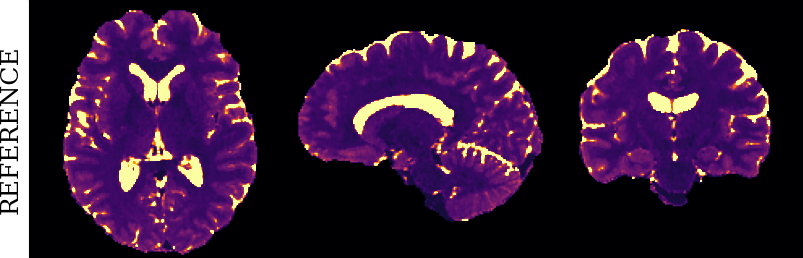} 
        \includegraphics[width=0.23\textwidth]{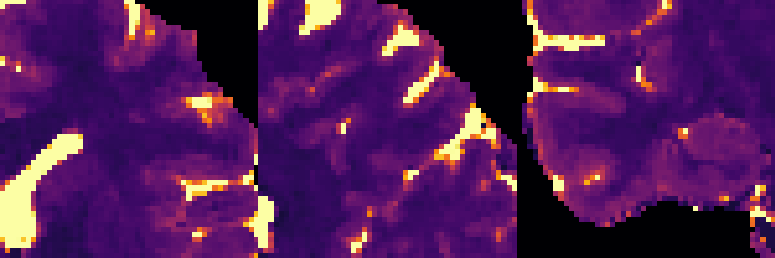} \\
        \includegraphics[width=0.24\textwidth]{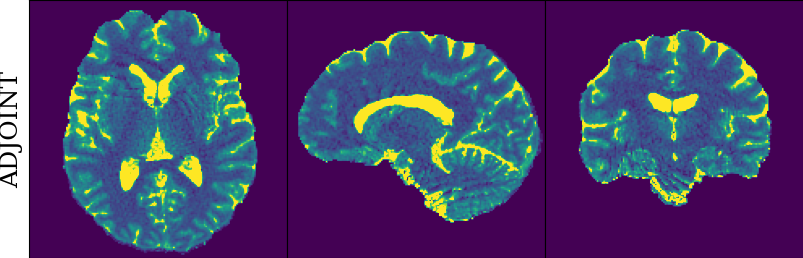}   
        \includegraphics[width=0.23\textwidth]{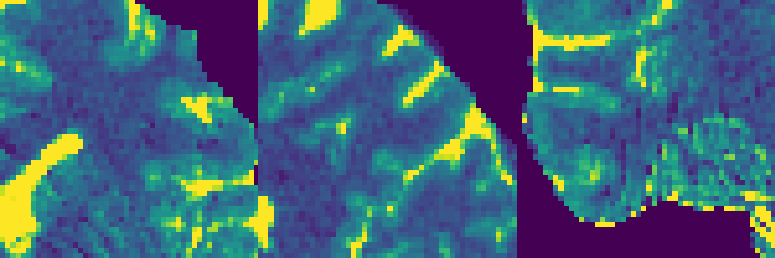} &
        \includegraphics[width=0.23\textwidth]{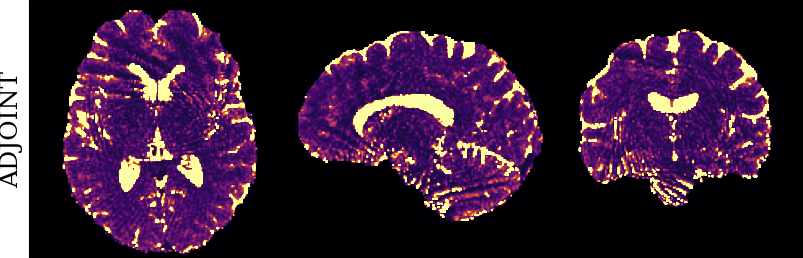} 
        \includegraphics[width=0.23\textwidth]{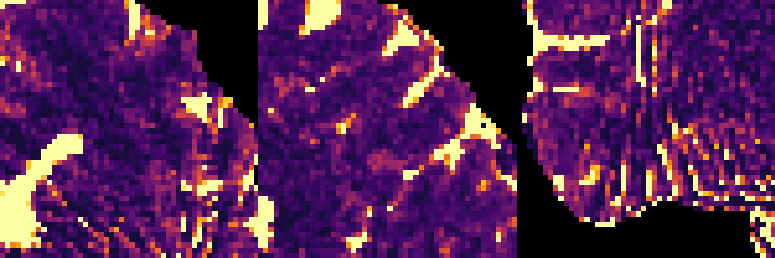} \\ 
        \includegraphics[width=0.24\textwidth]{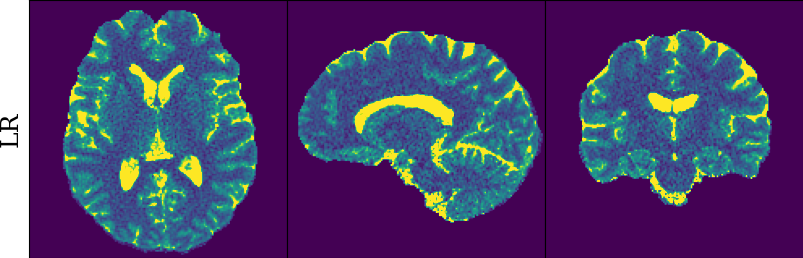} 
        \includegraphics[width=0.23\textwidth]{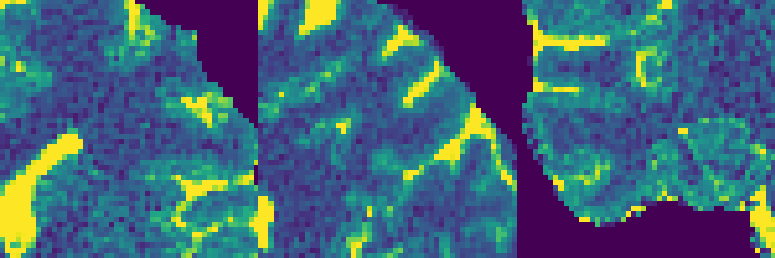} &
        \includegraphics[width=0.23\textwidth]{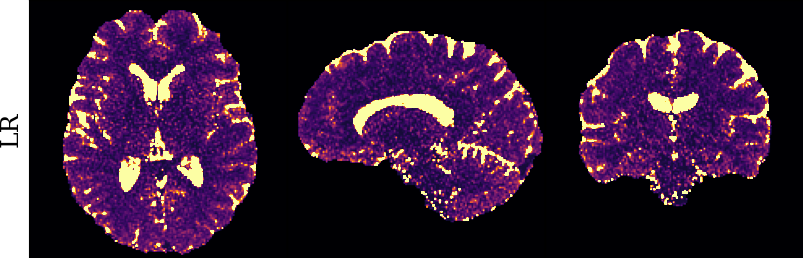}    
        \includegraphics[width=0.23\textwidth]{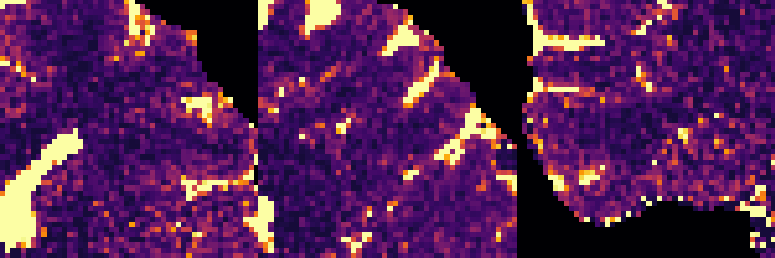} \\
        \includegraphics[width=0.24\textwidth]{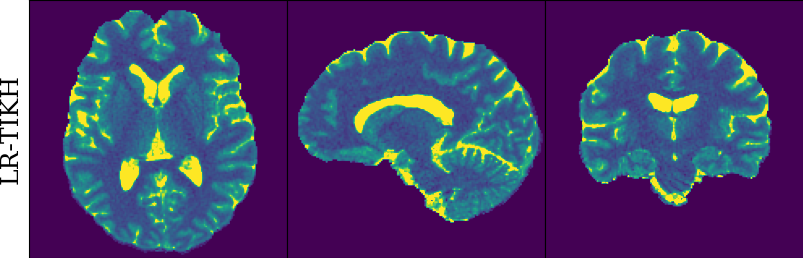} 
        \includegraphics[width=0.23\textwidth]{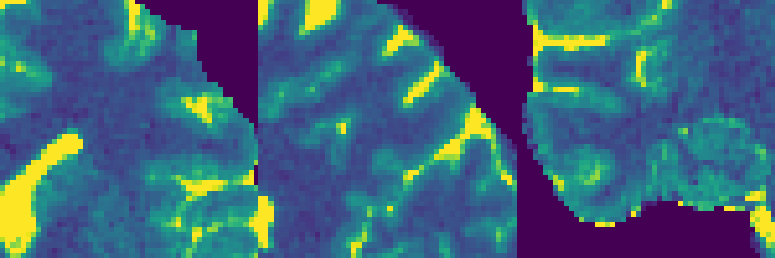} &
        \includegraphics[width=0.23\textwidth]{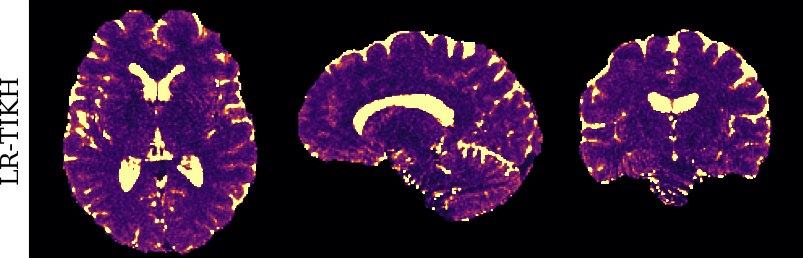}    
        \includegraphics[width=0.23\textwidth]{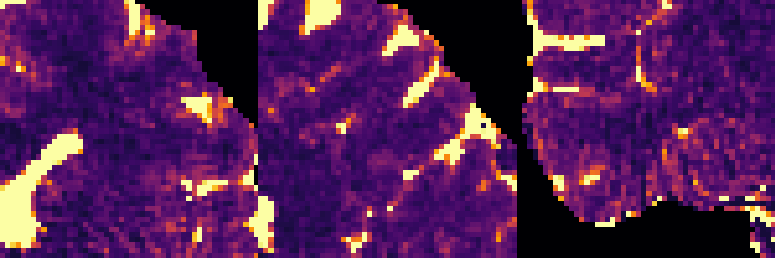} \\
        \includegraphics[width=0.24\textwidth]{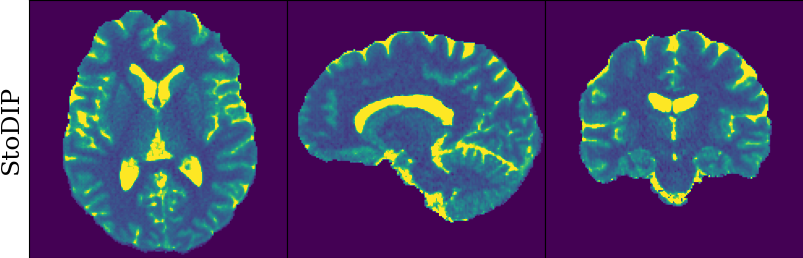} 
        \includegraphics[width=0.23\textwidth]{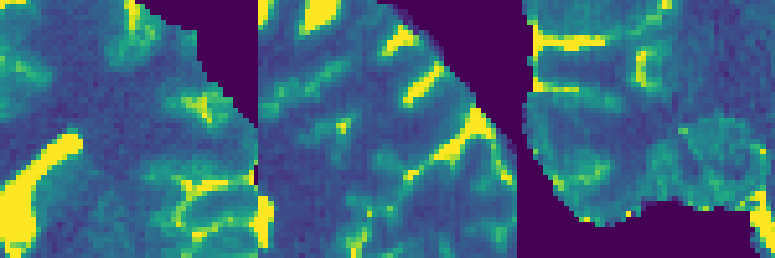} &
        \includegraphics[width=0.23\textwidth]{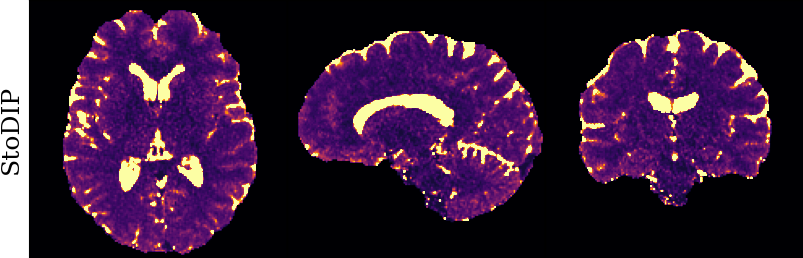} 
        \includegraphics[width=0.23\textwidth]{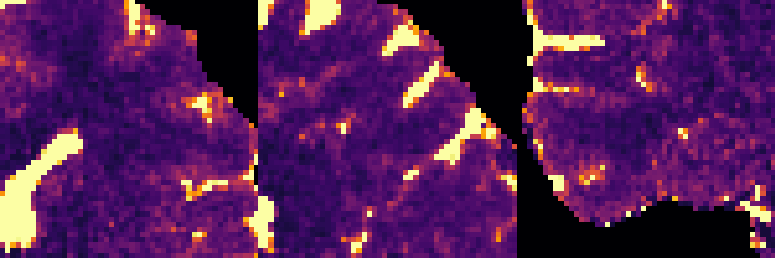} \\
        \includegraphics[width=0.24\textwidth]{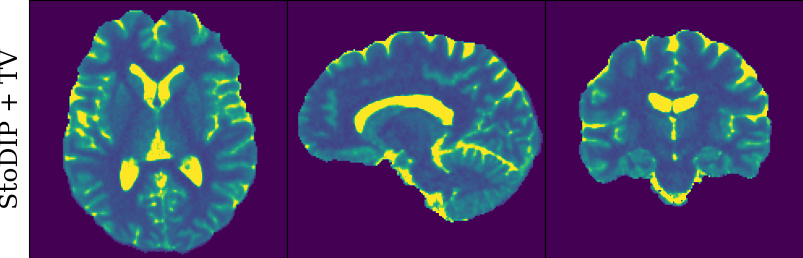} 
        \includegraphics[width=0.23\textwidth]{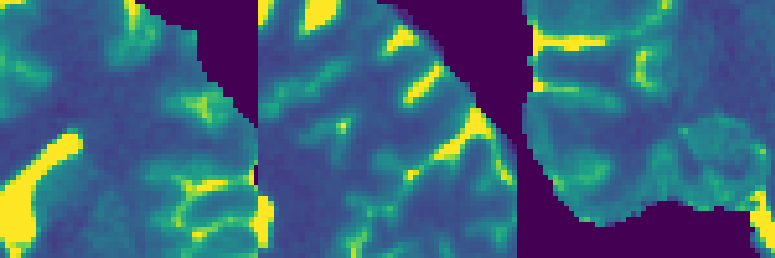} &
        \includegraphics[width=0.23\textwidth]{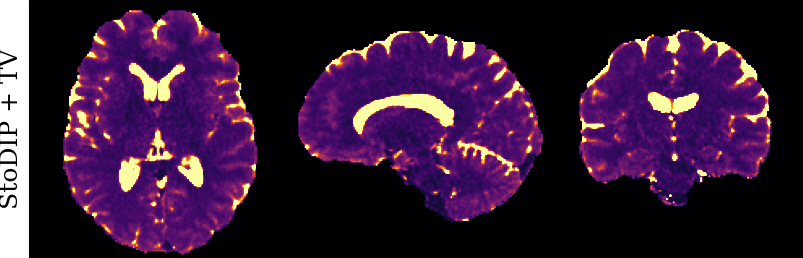} 
        \includegraphics[width=0.23\textwidth]{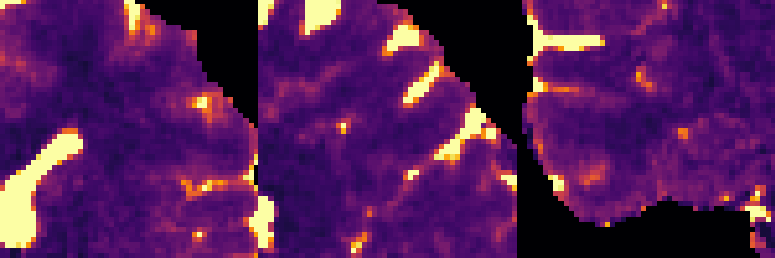} \\
        \includegraphics[width=0.5\textwidth]{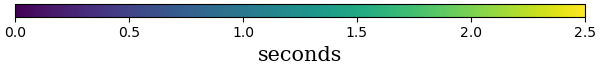} & 
        \includegraphics[width=0.5\textwidth]{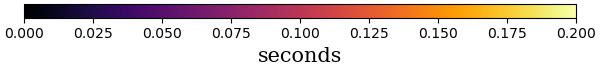} 
    \end{tabular}
    \caption{Reconstructed T1 and T2 maps for the assessed approaches on data with (retrospective) scan-time acceleration factor of 2. All maps have been masked using BET\cite{smith2002bet}. Right panels show zoomed in regions (electronic zoom recommended). }
    \label{fig:qmaps}
\end{figure}

The training and validation loss (Fig. ~\ref{fig:mc_loss_learning_rates}) served as valuable guides for architecture and training design. To demonstrate this, we selected a subset of experiments in which we kept certain variables fixed and only modified the one under inspection. From Fig. \ref{fig:mc_loss_learning_rates} (a)-(d), MAPEs curves show there is a clear setting with the preferred performance. Specifically, from Fig. \ref{fig:mc_loss_learning_rates}a, it is evident that using DRUNet yields a clear improvement over the original architecture. This could be attributed to the utilisation of residual units in addition to other architecture differences, such as the choice of upsampling operator (transpose convolution for DRUNet and trilinear for DIP). The choice of input (Fig. ~\ref{fig:mc_loss_learning_rates}b) demonstrated a consistent pattern among the tested options, with low rank (conjugate gradient) initialiser marginally outperforming the others. Fig. ~\ref{fig:mc_loss_learning_rates}cshows that by epoch 100, stochastic approaches exhibit lower reconstruction errors compared to non-stochastic methods. For example, adaptive LR with stochastic training yields a combined MAPE of 61.40\% (17.09\% T1 and 44.31\% T2) vs. 213.69\% (35.50\% T1 and 178.19\% T2) for the non-stochastic method. This fast convergence is due to adaptive LR and stochastic updates across coils. Despite similar execution times (~34 min for stochastic with adaptive LR vs. ~32 min for non-stochastic with fixed LR), the rapid error reduction makes StoDIP more attractive for training.. 
To assess the early stopping of DIP models, a key element, we present Fig. ~\ref{fig:mc_loss_learning_rates}d. The original work by Ulyanov et al. ~\cite{ulyanov2018deiporiginal} used 2k iterations, while Hamilton et al. ~\cite{hamilton2022dipmrf} used 30k. In contrast, our settings show StoDIP achieving competitive performance in under 500 epochs (4k iterations) on the entire volume. However, StoDIP can overfit to k-space measurements, affecting reconstruction accuracy, and thus the iteration at which it stops could have a greater impact. We show that adding a spatial penalty term addresses overfitting and instabilities. This is supported by the maps in Fig.\ref{fig:qmaps}, supplementary material Figures 1-3, and metrics in Table ~\ref{tab:results}. StoDIP reconstructions lack aliasing artifacts, and the TV regularizer (StoDIP + TV) further improves Q-Map reconstruction by reducing checkerboard artifacts observed in StoDIP outputs alone.

\section{Conclusion}
\label{sec:conclusion}
In this work, we introduce StoDIP, a pipeline capable of processing 3D-MRF data with a DIP-based model and within current GPU capabilities while achieving competitive performance. 
The strength of StoDIP relies on stochastic iterations across the coils, careful backbone network architecture design and the incorporation of efficient NUFFT libraries which enabled memory-efficiency and fast DIP convergence on challenging volumetric MRF data. This is achieved with only k-space measurements available, i.e., a ground-truth free reconstruction. StoDIP performance can be further improved by the addition of a spatial regularisation term on the reconstructed TSMI.  
We hypothesise that accuracy can be further enhanced by incorporating additional constraints such as Bloch consistency loss\cite{hamilton2022dipmrf}, which enforces the reconstructions to obey the physics of the problem. To improve each scan's reconstruction, a future avenue could benefit from a collection of scans, rather than a single one, to enrich the model capability with more knowledge of the problem.

\begin{credits}
\subsubsection{\ackname} 
This work has been carried out under the EPSRC grant\hspace{\fill}\linebreak EP/X001091/1. MC was funded by the INFN-CSN5 PREDATOR project (“Grant Giovani”). MT received support from the Italian Ministry of Health under the grant RC-L4 to IRCCS Fondazione Stella Maris

\subsubsection{\discintname}
The authors have no competing interests to declare that are
relevant to the content of this article.
\end{credits}
%
%
%
\bibliographystyle{splncs04}
\bibliography{references}
\end{document}